\newcommand{\ben}{\begin{enumerate}}
\newcommand{\een}{\end{enumerate}}
\newcommand{\beq}{\begin{equation}}
\newcommand{\eeq}{\end{equation}}
\newcommand{\beqn}{\begin{eqnarray}}
\newcommand{\eeqn}{\end{eqnarray}}

\newcommand{\beqd}{\begin{eqnarray*}}
\newcommand{\eeqd}{\end{eqnarray*}}
\newcommand{\bea}{\begin{array}}
\newcommand{\eea}{\end{array}}
\newcommand{\bcen}{\begin{center}}
\newcommand{\ecen}{\end{center}}
\newcommand{\btab}{\begin{tabular}}
\newcommand{\etab}{\end{tabular}}
\newcommand{\bsub}{\begin{subequations}}
\newcommand{\esub}{\end{subequations}}

\newcommand{\dfrac}{\displaystyle \frac}
\renewcommand{\vec}[1]{\mbox{\boldmath$#1$}}

\newcommand{\ls}{\left[}
\newcommand{\rs}{\right]}

\newcommand{\re}{\nonumber\\}
\documentstyle[twocolumn,epsf,rotate,aps]{revtex}
\begin{document}
\draft
\title{Pseudospin symmetry and its approximation in real nuclei}
\author{T.S. Chen, H.F. L\"{u}, J. Meng\thanks{e-mail: mengj@pku.edu.cn}, and S.-G. Zhou}
 \address{School of Physics, Peking University, Beijing 100871, China  \\
       Center for Theoretical Nuclear Physics, National Laboratory for
              Heavy Ion Physics, Lanzhou 730000, China \\
        Institute of Theoretical Physics, Chinese Academy of
              Sciences, Beijing 100080, China}
\date{\today}
\maketitle
\begin{abstract}
The origin of pseudospin symmetry and its broken in real nuclei
are discussed in the relativistic mean field theory. In the exact
pseudospin symmetry, even the usual intruder orbits have
degenerate partners. In real nuclei, pseudospin symmetry is
approximate, and the partners of the usual intruder orbits will
disappear.  The difference is mainly due to the pseudo spin-orbit
potential and the transition between them is discussed in details.
The contribution of pseudospin-orbit potential for intruder orbits
is quite large, compared with that for pseudospin doublets.  The
disappearance of the pseudospin partner for the intruder orbit can
be understood from the properties of its wave function.
\end{abstract}
\par

\narrowtext

%%%%%%%%%%%%%%%%%%%%%%%%%%%%%%%%%%%%
%     Introduction                 %
%%%%%%%%%%%%%%%%%%%%%%%%%%%%%%%%%%%%

%\section{INTRODUCTION}
Pseudospin symmetry was originally observed in spherical atomic
nuclei more than 30 years ago \cite{AHS69,HA69}, and later proved
to be a good approximation in deformed nuclei \cite{RR.73},
including triaxial deformed nuclei \cite{BBDB.97}. This symmetry
has been used to explain features of deformed nuclei, including
superdeformation \cite{Dudek.87}, and identical bands \cite
{Naza.90,Mot.91,ZJW.91} as well. The concept of pseudospin is
based on experimental observations that single particle states
with non-relativistic quantum numbers $(n,l,j=l+1/2)$ and
$(n-1,l+2,j=l+3/2)$ lie very close in energy and can therefore be
relabelled as pseudospin doublets:
 ($\tilde{n}=n-1$, $\tilde{\ell}=\ell+1$, $\tilde j=\tilde \ell \pm 1/2$).
  For example, the pair of orbits
($3s_{1/2},2d_{3/2}$) can be viewed as a pseudospin doublet
($2\tilde{p}_{1/2},2\tilde{p}_{3/2}$). While, all $p_{1/2}$ states
are pseudospin singlets $\tilde{s}_{1/2}$, and states with $n=1,
j=\ell+1/2$ are intruder orbital states without pseudospin
partner.

Since the suggestion of pseudospin symmetry in atomic nuclei,
there have been comprehensive efforts to understand its origin.
Apart from the rather formal relabeling of quantum numbers,
various proposals for an explicit transformation from the normal
scheme to the pseudospin scheme have been made in the last twenty
years, and the pseudospin symmetry is proved to be connected with
the special ratio in the strengths of the spin-orbit and
orbit-orbit interactions\cite{BHM.82,BDM.92,CMQ.92,BBD.96}.
Despite the long history of pseudospin symmetry, the origin of
symmetry has eluded explanation.

The relation between pseudospin symmetry and the relativistic mean
field (RMF) theory \cite{SW.86} was first noted in \cite{BDM.92},
in which Bahri et al.found that the RMF theory explains
approximately the special ratio of the strengths of spin-orbit and
orbit-orbit interactions in the non-relativistic calculations for
exact pseudospin symmetry. No great progress was achieved until
Ginocchio revealed that the pseudo-orbital angular momentum is
nothing but the ``orbital angular momentum" of the lower component
of the Dirac spinor and built the connection between pseudospin
symmetry and the equality in magnitude but difference in sign of
the scalar potential $V_s(r)$ and vector potential $V_v(r)$
\cite{Gi97,GL.98}. Based on Relativistic Continuum
Hartree-Bogoliubov (RCHB) \cite{ME.98,MR.96} , it is shown that
pseudospin symmetry is exact under a more general condition,
$d(V_s+V_v)/dr=0$, and the quality of the pseudospin approximation
in real nuclei is connected with the competition between the
pseudo-centrifugal barrier and the pseudospin-orbital potential
\cite{Meng98,Meng99}.

Here, in this paper, we will show why the pseudospin symmetry is
approximate and broken in real nuclei, and explain the partner of
intruder orbit is missing. We will first solve Dirac equation for
two kinds of harmonic oscillator potentials with exact spin
symmetry and pseudospin symmetry, respectively. Taking $^{208}$Pb
as an example, we will investigate how the pseudospin symmetry is
a good approximation for normal orbits and why intruder orbit
fails to have partner in real nuclei. A short summary is given at
the end.

%%%%%%%%%%%%%%%%%%%%%%%%%%%%%%%%%%%%
%     FORMULISM                    %
%%%%%%%%%%%%%%%%%%%%%%%%%%%%%%%%%%%%
%\section{TWO SU(2) SYMMETRIES IN DIRAC EQUATION}
%\subsection{FORMULISM OF DIRAC EQUATION}

The Dirac equation of a nucleon with mass $M$ moving in an
attractive scalar potential $V_s({\vec r})$ and a repulsive vector
potential $V_v({\vec r})$ can be written as,
 \beq
 \label{eq:dirac}
 \ls\;\vec \alpha\cdot \hat {\vec p}
  +\beta(M+V_s) +V_v\rs \Psi_i\; =\; E_i \Psi_i.
 \eeq
For spherical nuclei, the nucleon angular momentum $\hat {\vec
J}$, and $\hat \kappa=-\hat \beta (\hat \sigma\cdot\hat{\vec L} +
1)$ commute with the Dirac Hamiltonian, where $\hat \beta$,
$\sigma$, and $\vec L$ are respectively the Dirac matrix, Pauli
matrix, and orbital angular momentum. The eigenvalues of $\hat
\kappa$ are $\kappa = \pm(j+1/2)$ with $-$ for aligned spin
($s_{1/2}$, $p_{3/2}$, etc.) and $+$ for unaligned spin
($p_{1/2}$, $d_{3/2}$, etc.).  For a given $\kappa=\pm 1,\pm 2,
...$, $j =|\kappa|-\frac 1 2$, $\ell = |\kappa+1/2|-\dfrac 1 2$ ,
$\tilde{\ell} = |\kappa-1/2|-1/2$. The wave functions can be
classified according to their angular momentum $j$, $\kappa$, and
the radial quantum number $n$. In terms of the upper and lower
radial functions $F_{n\kappa}(r)$ and $G_{n\kappa}(r)$,  Eq.
(\ref{eq:dirac}), becomes\cite{Greiner}:
 \beqn
 (\dfrac d {dr} + \dfrac \kappa r )\; F_{n\kappa}(r) &=& (M + E_{n\kappa} -\Delta)\;
  G_{n\kappa}(r)
  \label{R-F},\\
 (\dfrac d {dr} - \dfrac \kappa r )\; G_{n\kappa}(r) &=& (M-E_{n\kappa} +V)\;
  F_{n\kappa}(r)
  \label{R-G},
 \eeqn
with $\Delta=V_v-V_s$ and $V=V_s+V_v$. Eliminating
$G_{n\kappa}(r)$ or $F_{n\kappa}(r)$ one can get the following
Schr\"{o}dinger-like equations,
 \beqn
  \label{eq:spin}
 \left[ \dfrac {d^2} {dr^2} - \dfrac {\ell(\ell+1)}{r^2} -(M+E_{n\kappa}-\Delta)(M-E_{n\kappa}+V)  \right. \re
 \left. + \dfrac {\displaystyle{\dfrac {d\Delta}{dr}\;(\dfrac d {dr}+\dfrac \kappa r) } }
  {M+E_{n\kappa} - \Delta}\right]F_{n\kappa}(r)=0,  \ \ \
  \eeqn
  \beqn
 \label{eq:pseudospin}
 \left[ \dfrac {d^2} {dr^2} - \dfrac {\tilde{\ell}(\tilde{\ell}+1)}{r^2}-(M+E_{n\kappa}-\Delta)(M-E_{n\kappa}+V)  \right. \re
  \left. - \dfrac {\displaystyle {\dfrac {dV}{dr}(\dfrac d {dr}-\dfrac \kappa r)}}
  {M-E_{n\kappa}+V}\right]G_{n\kappa}(r)=0, \ \ \
 \eeqn
 and the same result can be obtained by solving either Eq.(\ref{eq:spin}) or
 Eq.(\ref{eq:pseudospin}).

%\subsection{SPIN SYMMETRY}

For $\Delta=0$ or $\displaystyle{\dfrac {d\Delta}{dr}}=0$,
Eq.(\ref{eq:spin}) is reduced to
 \beqn
 \label{eq:Fell}
 & & \left[\dfrac {d^2}{dr^2} - \dfrac {\ell(\ell+1)}{r^2} \right]\; F_{n\kappa}(r) \re
 &=& \left[(M+E_{n\kappa}-\Delta)(M-E_{n\kappa}+V) \right]\; F_{n\kappa}(r).
 \eeqn
The eigen energies, $E_{n\kappa}$, depend only on $n$ and $\ell$,
i.e., $E_{n,\kappa} =E(n, \ell(\ell+1))$ . For $\ell \ne 0$, the
states with $j=\ell \pm 1/2$ are degenerate. This is a SU(2) spin
symmetry.

For the Dirac equation with the following vector and scalar
potentials,
 \beq
 V_v(r) = V_s(r)=\dfrac 1 4 M \omega ^2 r^2 \ ,
 \eeq
$\Delta=V_v(r)-V_s(r)=0$, Eq.(\ref{eq:Fell}) can be solved
analytically with the eigen energy,
 \beqn
 \label{eq:harmonic2}
&&(E_{n\kappa}-M)\sqrt{\dfrac {E_{n\kappa}+M}{2M} } \re
% &=& \;\omega \; (2n'+\ell + \dfrac 3 2) ,\ (n'=0,1,...),\ \re
&=& \;\omega \; (2n+\ell - \dfrac 1 2) \ \ \ (n=1,2,...).
 \eeqn
Now states with the same $\it n$ and $\ell$ will be degenerate. Of
course as the potential is a special harmonic oscillator, the
spectrum is more degenerated. The corresponding upper radial
wavefunction is,
 \beqn
 \label{eq:upper}
  F_{n\kappa}(r) = \lambda_1 \;
  (\alpha r)^{\ell + 1}\;e^{-\dfrac 1 2 \alpha^2 r^2} \;
  L_{n-1}^{\ell+1/2}(\alpha^2  r^2) , \re
  (\  \alpha^2 = \omega \sqrt{M(E_{n\kappa}-M)}\ ), \ \
\eeqn
 where $\lambda_1$ is a normalization constant and
$L^{\ell+1/2}_{n-1}$ is the Laguerre polynomials. The lower radial
wave function can be obtained from Eq.(\ref{R-F}). For the spin
partners, although the upper wave functions are the same, their
lower wave functions are different because their $\kappa$ are
different.

In Fig. \ref{fig:sd-broken}, the energy spectrum for the Dirac
equation with $V_v=V_s=2.5\; r^2$ is given by solving
Eq.(\ref{eq:Fell}). Of course, $\Delta=0$ or $\displaystyle{\dfrac
{d\Delta}{dr}}=0$ means the missing of the spin-orbital
interaction, which cannot be true for real nuclei.

%\subsection{PSEUDOSPIN SYMMETRY}

For $V=0$ or $ \displaystyle{\dfrac {dV}{dr}}\; =0$, Eq.(
\ref{eq:pseudospin}) is reduced as
 \beqn
 \label{eq:kl}
  & & \ls \dfrac {d^2}{dr^2} - \dfrac {\tilde{\ell}(\tilde{\ell}+1)}{r^2} \rs\; G_{n\kappa}(r) \re
  &=& \ls\; (M+E_{n\kappa}-\Delta)(M-E_{n\kappa}+V)  \; \rs\; G_{n\kappa}(r).
 \eeqn
The solution of the Dirac equation can be obtained from
Eq.(\ref{eq:kl}) and its eigen energies, $E_{n\kappa}$, depend
only on $n$ and $\tilde{\ell}$, i.e., $E_{n\kappa} =
E(n,\;\tilde{\ell}(\tilde{\ell}+1)\;)$.  For $\tilde{\ell}\ne 0$,
the states with $j=\tilde\ell \pm 1/2$ are degenerate. This is the
pseudospin SU(2) symmetry. Except pseudospin singlet states with
$\tilde{\ell}=0$, every aligned state $(\;
j=\ell+1/2=\tilde{\ell}-1/2 \; )$ will have a degenerate unaligned
partner $(\; j=\ell-1/2=\tilde{\ell}+1/2 \;)$, and vice versa.
Even the usual intruder orbital states may have partner now. These
characters are very similar to the spin SU(2) symmetry.

For the Dirac equation with the following vector and scalar
potentials,
 \beq
 V_v(r) = -V_s(r) =\dfrac 1 4 M \omega ^2 r^2 \ ,
 \eeq
$V=V_v(r)+V_s(r) =0$, Eq.(\ref{eq:kl}) can be solved analytically
with the eigen energy,
 \beqn
 \label{eq:harmonic}
 & &(E_{n\kappa}+M)\sqrt{\dfrac {E_{n\kappa}-M} {2M} } \re
 % &=& \omega \;(2n'+\tilde{\ell}+ \dfrac 3 2)\ \ \ (n'=0,1,...), \re
 &=& \omega \;(2n+\tilde{\ell}- \dfrac 1 2) \ \ \ (n=1,2,...).
 \eeqn
The states with the same $\it n$ and $\tilde{\ell}$ will be
degenerate. The corresponding lower radial wavefunction is,
 \beqn
G_{n\kappa}(r) = \lambda_2 \; (\alpha
r)^{\tilde{\ell}+1}\;e^{-\dfrac 1 2 \alpha^2 r^2}
 L_{n-1}^{\tilde{\ell}+\dfrac 1 2}(\alpha^2 r^2) \, \re
(\  \alpha^2 = \omega \sqrt{M(E_{n\kappa}+M)}\  ), \ \
 \eeqn
where $\lambda_2$ is the normalization constant and
$L^{\tilde{\ell}+1/2}_{n-1}$ is the Laguerre polynomials. The
upper radial wave function can be obtained from Eq.(\ref{R-G}).
For the pseudospin partners, although the lower wave functions are
the same, their upper wave functions are different due to
$\kappa$.

In Fig.\ref{fig:sd-complete}, the energy spectrum for the Dirac
equation with $V_v=-V_s=2.5\; r^2$ is given by solving
Eq.(\ref{eq:kl}). Similar to Fig.\ref{fig:sd-broken}, the spectrum
is more degenerate due to the special harmonic oscillator used
here. To make sure that the radial quantum number in
Fig.\ref{fig:sd-complete} is correct and even the intruder orbits
have pseudospin partners, the spectrum in
Fig.\ref{fig:sd-complete} is confirmed by solving the Dirac
equation (\ref{eq:dirac}) numerically as in Refs.
\cite{ME.98,MR.96}.

%\section{PSEUDOSPIN SYMMETRY IN REAL NUCLEI}

Now the questions arise: in real nuclei why the pseudospin
symmetry is broken and there is no partner for intruder orbit.

In real nuclei, $\dfrac {dV} {dr}\ne 0$, and pseudospin symmetry
is only an approximation. The quality of the pseudospin symmetry
approximation depends on the competition between the contributions
of the pseudo-centrifugal potential, $\dfrac
{\tilde{\ell}(\tilde{\ell}+1)} {r^2}$, and the pseudospin-orbit
potential, $\dfrac {dV} {dr} \dfrac {\kappa} {r(M-E+V)}$, in Eq.
(\ref{eq:pseudospin}).

In \cite{Meng98,Meng99}, Meng et al.analyzed the competition of
the two potentials for pseudospin partners, and found that the
contribution of the pseudospin-orbit potential is small, compared
with that of the pseudo-centrifugal potential. Following the
arguments and conjectures in Ref. \cite{Meng98,Meng99}, pseudospin
symmetry is discussed for the nuclei with deformed shape or
isospin asymmetry \cite{STA.98,STA.00,AFM.01}.

To understand why the intruder orbit has no partner is a
challenging problem. Lots of works have been done along this line.
In Ref.\cite{Gino.01.PRL,Gino.01.PLB}, the wave functions of the
pseudospin partners and the structure of the radial nodes have
been investigated. Here combining the energy spectrum in
Fig.\ref{fig:sd-complete} with that obtained from RCHB
\cite{ME.98,MR.96}, we will try to understand why the intruder
orbit has no partner.

In Fig.\ref{fig:intruder}, multiplied with the wave function
$G^2$, the effective pseudospin-orbit potential $\dfrac \kappa r
\dfrac {dV}{dr}$ and the effective centrifugal barrier $(M-E+V)
\dfrac {\tilde{\ell}(\tilde{\ell}+1)}{r^2}$ in $^{208}$Pb have
been compared in detail for the pseudospin singlets, the
pseudospin doublets, and the intruder orbits.

For pseudospin singlets, we choose $1p_{1/2}$ and $2p_{1/2}$ as
examples, and find that the pseudo-centrifugal potential is always
zero. For the usual pseudospin doublets, the contribution of the
effective pseudospin-orbit potential is much smaller than that of
the effective pseudo-centrifugal potential, and the pseudospin
symmetry is a good approximation. For the intruder orbits, the
contribution of the effective pseudospin-orbit potential is
comparable with and sometimes even larger than that of the
effective pseudo-centrifugal potential. It can be seen that the
pseudospin-orbit potential is responsible for the broken of
pseudospin symmetry. So far it is not explained where the
pseudospin partner of the intruder orbit in real nuclei has gone.
In the following we will try to get the answer from the properties
of the wave function.

For bound states, their solution can be obtained from either one
of the Schr\"{o}dinger-like equations, Eqs.(\ref{eq:spin}) and
(\ref{eq:pseudospin}). Let $F(r)=r^{-\kappa}\; f(r)$, from
Eq.(\ref{R-F}) one get:

\beq
  \dfrac {df}{dr}=r^{\kappa} (M+E_{n\kappa}-\Delta)\;G .
  \label{eq:df}
\eeq
   For $r \sim 0$, the solution of bound states for
Eqs.(\ref{R-F}) and (\ref{R-G}) can be obtained analytically as:

\beqn
   F(r) &=& \lambda\; {\rm Sign}(\kappa)\; \sqrt{\dfrac {E_{n\kappa}+M-\Delta}{2M+V-\Delta}} \sqrt{r} J_{|\kappa+1/2|}(k\,r) \\
   \label{eq:bf}
   G(r) &=&  \lambda\; \sqrt{\dfrac {E_{n\kappa} -M -V}{2M+V-\Delta}} \sqrt{r} J_{|\kappa-1/2|}(k\,r),
    \label{eq:bg}
\eeqn
 where $k=\sqrt{-(M+E-\Delta)(M-E+V)}$ and $\lambda$ is a normalization factor ( here assumed to be
 positive ). For $\kappa>0$, if the
node number  of lower wave function $G$ is zero: $n_G=0$, then $G$
will be always positive. As $r \sim 0$, one has:

\beq
     J_n(r)= r^{n}\; ( \dfrac {2^{-n}} {\Gamma(1+n)} + O[r]^2 ),
      \label{eq:J0}
\eeq
 and depending on $\kappa$, $f(r)$ can be written as:

\beq \displaystyle
     f(r)\sim \lambda \sqrt{\dfrac{E_{n\kappa}+M-\Delta}{2M+V-\Delta}}\;\dfrac {2^{-|\kappa+1/2|}}{\Gamma(1+|\kappa+1/2|)}
    \times \left\{
    \begin{array}{cc}
          r^{2\kappa+1}  & \kappa>0 \\
      -1  & \kappa<0
    \end{array}
          \right.
     \label{eq:fk}
\eeq
 As the factor $M+E_{n\kappa}-\Delta$ in Eq.(\ref{eq:df}) is
 always positive, for  $n_G=0$ and $\kappa>0$,
$\dfrac {df}{dr}$ will be always positive, i.e., $f(r)$ increases
with $r$. While for bound states, the wave function at large $r$
should decrease as an exponential function of $r$, i.e.,
$f(\infty)=0$. Therefore for $n_G = 0$ and  $\kappa>0$, the
boundary condition of $f$ for bound state can not be satisfied and
there is no corresponding bound states. This means that the
pseudospin partner of the intruder orbit will disappear. Similar
conclusion has been drawn in Ref. \cite{Gino.01.PLB,Gino.01.PRL}.
To restore the pseudospin symmetry for intruder orbit, one
possibility is that the factor $(M+E_{n\kappa}-\Delta)$ can change
sign with $r$, as the case in Fig.\ref{fig:sd-complete}.

%\section{SUMMARY}

In summary, the origin of pseudospin symmetry and its broken in
real nuclei are discussed in the RMF theory. In the exact
pseudospin symmetry, all the orbits except the singlets p$_{1/2}$
have partners. In real nuclei, pseudospin symmetry is approximate,
and the partners of the usual intruder orbits will disappear. The
competition between the pseudo-centrifugal and pseudospin-orbit
potentials decides the quality of pseudospin symmetry. The
contribution of pseudospin-orbit potential for intruder orbits is
quite large, compared with that for pseudospin doublets.  The
disappearance of the pseudospin partner for the intruder orbit can
be understood from the properties of its wave function.

 Valuable discussions with Prof. J.Y. Zeng are gratefully acknowledged.
 This work was partly
supported by the Major State Basic Research Development Program
Under Contract Number G2000077407 and the National Natural Science
Foundation of China under Grant No. 10025522, 10047001, and
19935030 .

%\newpage

\begin{figure}
%{\centerline{\epsfxsize 6cm \epsffile{Fig1.eps}}}
\vspace*{20pt}
   \caption{The eigen energy for the Dirac equation with $V_v(r) =
V_s(r)=\dfrac 1 4 M \omega ^2 r^2 $, $M=10.0$, and $\omega=1.0$. }
   \label{fig:sd-broken}
\end{figure}

\begin{figure}
%{\centerline{\epsfxsize 6cm \epsffile{Fig2.eps}}}
\vspace*{20pt}
 \caption{Same as Fig.\ref{fig:sd-broken} but for $V_v(r) =- V_s(r)= \dfrac 1
4 M \omega ^2 r^2 $. } \label{fig:sd-complete}
\end{figure}

\begin{figure}
  %{\centerline{\epsfxsize 6cm \epsffile{Fig3a.eps}}}
  \centering{(a) Pseudospin singlet states}
\\[10pt]
  %{\centerline{\epsfxsize 6cm \epsffile{Fig3b.eps}}}
    \centering{(b) Pseudospin doublet states}
\\[10pt]
   %{\centerline{\epsfxsize 6cm \epsffile{Fig3c.eps}}}
    \centering{(c) Intruder orbits}
  \\[10pt]
\vspace*{20pt}
 \caption{The competition between the effective pseudo-centrifual
$G^2(r) (E-M-V) \tilde{\ell}(\tilde{\ell}+1) / r^2$ ( solid line )
and pseudospin-orbital potentials $-G^2(r) \dfrac {dV}{dr}\;\dfrac
{\kappa}{r} $ ( dashed line ). } \label{fig:intruder}
\end{figure}

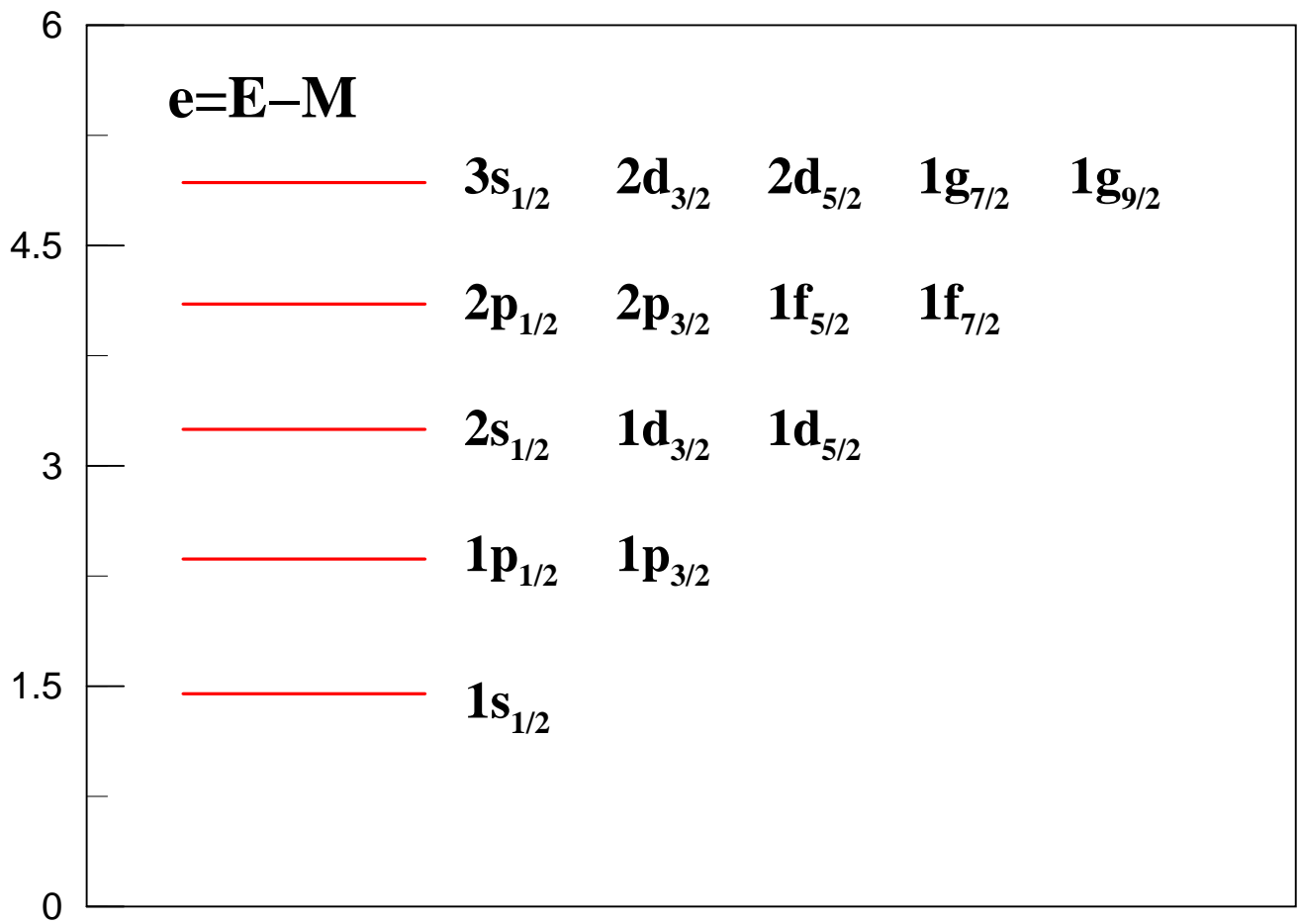
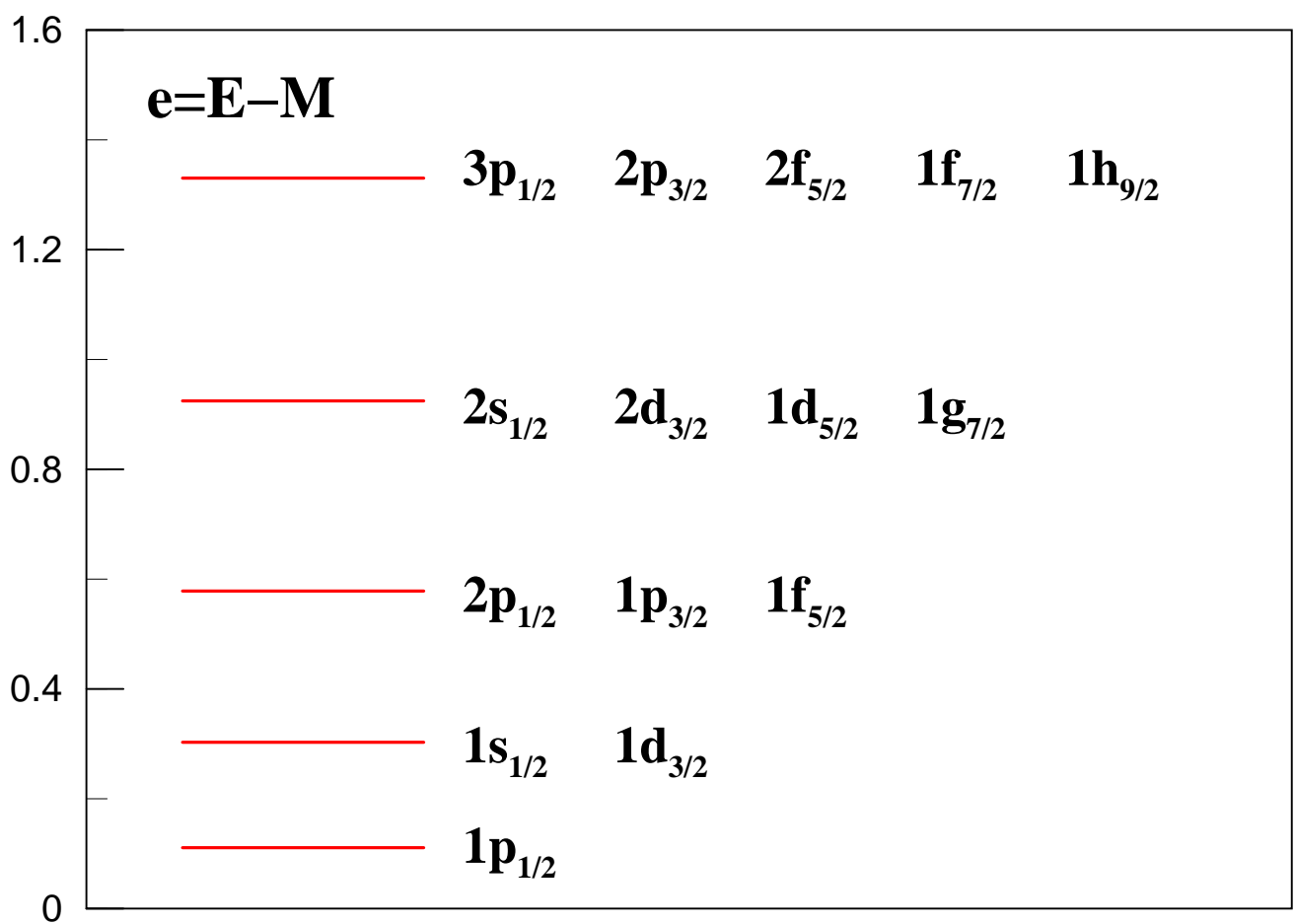
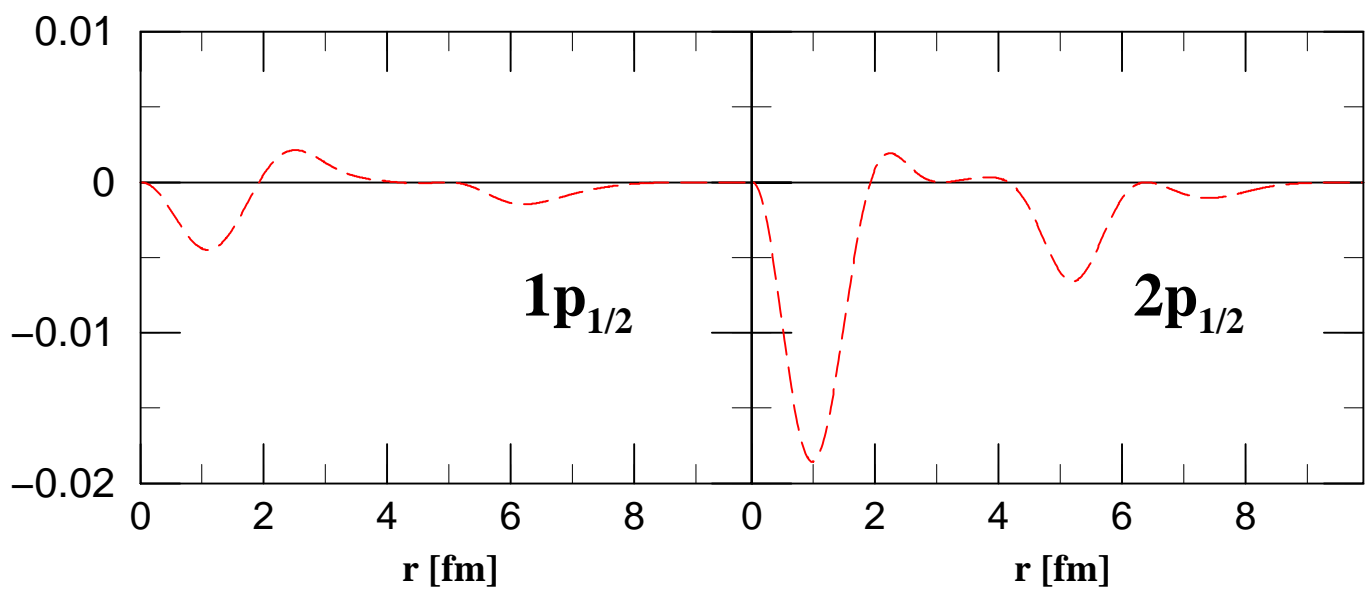
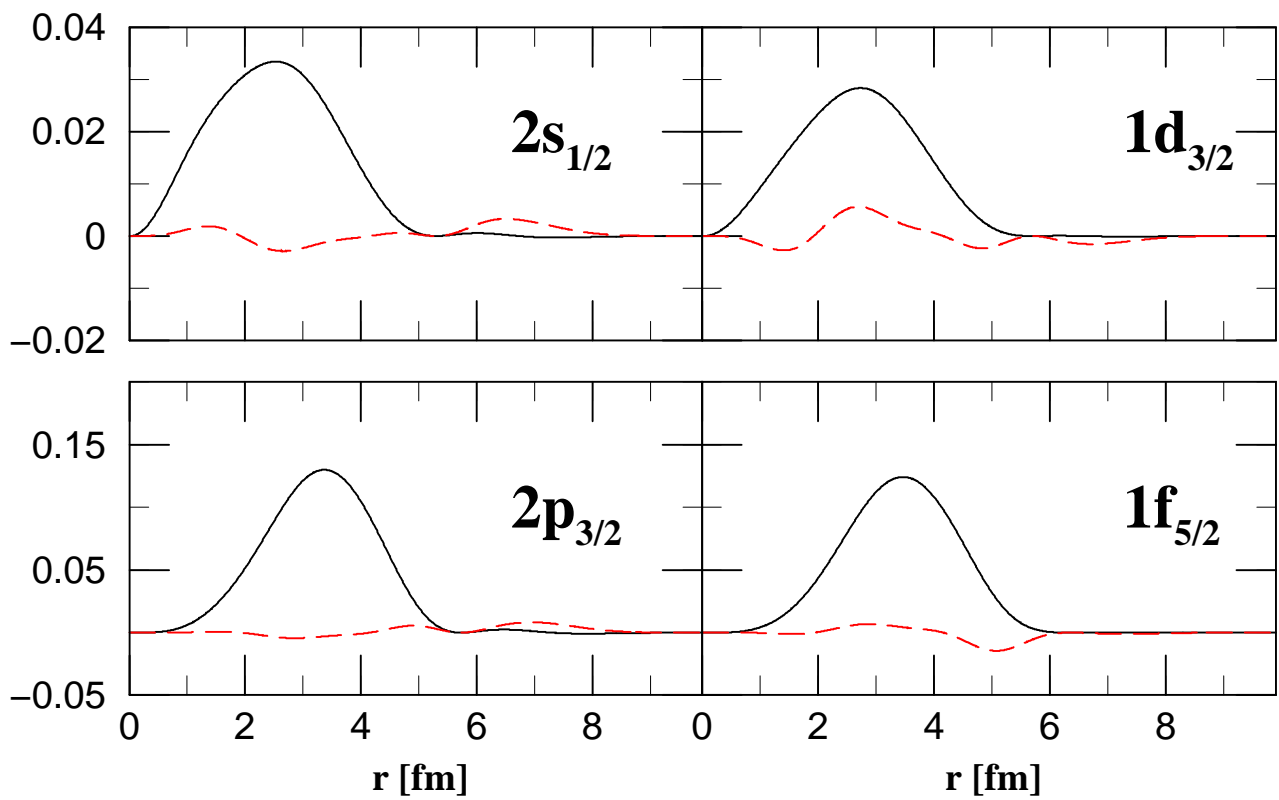
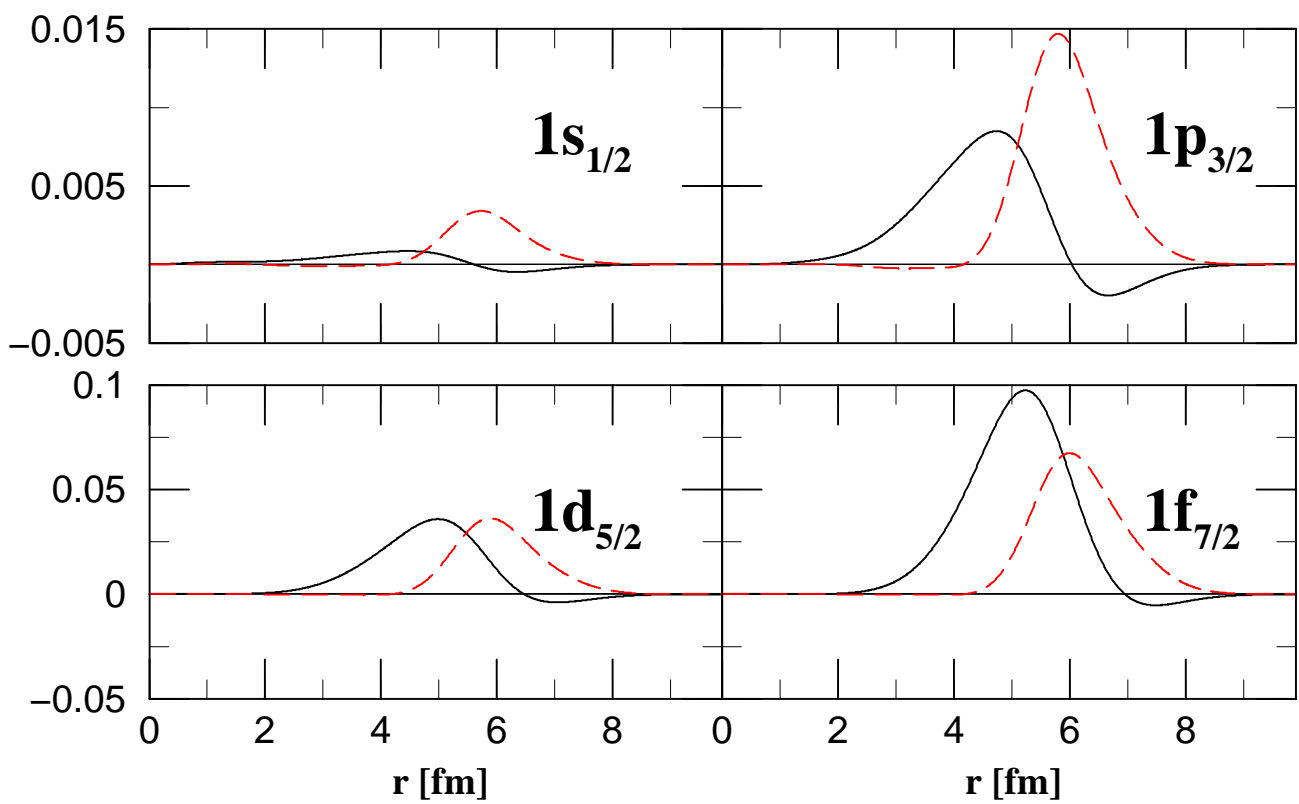
\end{document}